# PREDICTION AND ANALYSIS OF TIDES AND TIDAL CURRENTS

## *Jian-Jun Shu*

School of Mechanical & Aerospace Engineering, Nanyang Technological University
50 Nanyang Avenue, Singapore 639798

**ABSTRACT**

An efficient algorithm of tidal harmonic analysis and prediction is presented in this paper. Some conditions are found by means of the known approximate relationships between the harmonic constants of the tidal constituents. A system of linear equations for least squares solutions under these restricted conditions is obtained. In the case of inadequate data, ill conditioning in the system of equations that has appeared in other algorithms is conveniently avoided. In solving the resultant normal equations, the Goertzel iteration is adopted so that the whole computation time is dramatically reduced.

**INTRODUCTION**

Tides and tidal currents offer clean and inexhaustible energy sources. Better prediction and analysis of tides and tidal currents are crucial to utilize hydro-dams more efficiently as energy generators. Tides are cyclic variations in the level of seas and oceans, while tidal currents are cyclic variations in the motion of seas and oceans. The present understanding of tides and tidal currents as natural phenomena due to the gravitational forces of the sun and moon acting on a rotating earth came from the development of Newton's gravitation theory [1]. Harmonic techniques were first used to analysis and predict tides and tidal currents by Thomson [2] and expended by Darwin [3], Harris [4] and Doodson [5]. Tides and tidal currents may be considered as the sum of tidal constituents according to harmonic analysis. With the development of digital computers least squares technique is used to evaluate the tidal constituents from observed data and this is a principal method used today.

The harmonic method of tidal analysis has been further refined for improvement in accuracy of tidal prediction. A method for superfine resolution of tidal harmonic constituents has been developed by Amin [6-8] adding a corrective step into the harmonic method. The species concordance method has been developed by George & Simon [9] and Simon [10] using relationships between species of the tide at the studied station and at a reference station where the tide is well known or easily predicted. Here we reexamine the harmonic method from a practical point of view and propose an efficient algorithm of tidal harmonic analysis and prediction.

**HARMONIC METHOD FOR REGULAR OBSERVATIONS**

Let us consider real-time regular observed data of tidal height $h_n = h(t_0 + n\Delta T)$ $(n = 0,1,\cdots,2N)$, where $t_0$ is the initial time, $2N+1$ is the number of the real-time observed data, and $\Delta T$ is the sampling time interval. The tidal height can be expressed as a sum of cosine functions plus random errors denoted by $r_n = r(t_0 + n\Delta T)$.

$$h_n = x_0 + \sum_{m=1}^{M} R_m \cos[\sigma_m(t_0 + n\Delta T) - \varphi_m] + r_n, \quad n = 0,1,\cdots,2N \quad (1)$$

where

$$R_m = f_m H_m, \quad \varphi_m = g_m - \chi_m, \quad m = 1,2,\cdots,M.$$

$f_m$, $H_m$, $\sigma_m$, $g_m$, and $\chi_m$ are the node factor, mean amplitude, angular velocity, epoch, and astronomical argument of the $m$th tidal constituent respectively. $M$ is the number of tidal constituents.

Eq. (1) can be rewritten as

$$h_n = x_0 + \sum_{m=1}^{M}\{x_m\cos[(n-N)\sigma_m\Delta T] + y_m\sin[(n-N)\sigma_m\Delta T]\} + r_n, \quad (2)$$
$$n = 0,1,\cdots,2N,$$

where

$$\{x_m, y_m\} = R_m\{\cos[\varphi_m - \sigma_m(t_0 + N\Delta T)], \sin[\varphi_m - \sigma_m(t_0 + N\Delta T)]\}$$
$$m = 1,2,\cdots,M.$$

Letting $\sigma_0 = 0$ and using matrix notation, Eq. (2) can be expressed as

$$R = H - QZ \quad (3)$$

where

$$R = \left(\frac{r_0}{\sqrt{2}}, r_1, \cdots, r_{2N-1}, \frac{r_{2N}}{\sqrt{2}}\right)^T, \quad H = \left(\frac{h_0}{\sqrt{2}}, h_1, \cdots, h_{2N-1}, \frac{h_{2N}}{\sqrt{2}}\right)^T.$$

$$Z^T = (X^T, Y^T),$$

and

$$X^T = (x_0, x_1, \cdots, x_M), \quad Y^T = (y_1, y_2, \cdots, y_M).$$

The column vectors of the coefficient matrix $Q$ are

$$Q_0, Q_1, \cdots, Q_{2M},$$

where

$$Q_m^T = (\frac{1}{\sqrt{2}}\cos[-N\sigma_m\Delta T], \cos[-(N-1)\sigma_m\Delta T], \cdots,$$
$$\cos[(N-1)\sigma_m\Delta T], \frac{1}{\sqrt{2}}\cos[N\sigma_m\Delta T]) \quad m = 0,1,\cdots,M,$$

$$Q_m^T = (\frac{1}{\sqrt{2}}\sin[-N\sigma_m\Delta T], \sin[-(N-1)\sigma_m\Delta T], \cdots,$$
$$\sin[(N-1)\sigma_m\Delta T], \frac{1}{\sqrt{2}}\sin[N\sigma_m\Delta T]), \quad m = M+1, M+2,\cdots,2M.$$

We make

$$\|R\|_2^2 = \|H - QZ\|_2^2 = \min,$$

where $\|R\|_2$ is an Euclidean norm of $R$. Then

$$\frac{\partial\|R\|_2^2}{\partial x_0} = \frac{\partial\|R\|_2^2}{\partial x_m} = \frac{\partial\|R\|_2^2}{\partial y_m} = 0, \quad m = 1,2,\cdots,M.$$

That is

$$Q^T Q Z = Q^T H. \quad (4)$$

We arrive at

$$Q^T Q = \begin{bmatrix} F & O \\ O & G \end{bmatrix}, \quad Q^T H = \begin{bmatrix} C \\ S \end{bmatrix},$$

where

$$F_{lm} = \frac{1}{2}\left\{\frac{\sin[N(\sigma_l - \sigma_m)\Delta T]}{\tan[\frac{1}{2}(\sigma_l - \sigma_m)\Delta T]} + \frac{\sin[N(\sigma_l + \sigma_m)\Delta T]}{\tan[\frac{1}{2}(\sigma_l + \sigma_m)\Delta T]}\right\}, \quad l,m = 0,1,\cdots,M,$$

$$G_{lm} = \frac{1}{2}\left\{\frac{\sin[N(\sigma_l - \sigma_m)\Delta T]}{\tan[\frac{1}{2}(\sigma_l - \sigma_m)\Delta T]} - \frac{\sin[N(\sigma_l + \sigma_m)\Delta T]}{\tan[\frac{1}{2}(\sigma_l + \sigma_m)\Delta T]}\right\}, \quad l,m = 1,2,\cdots,M.$$

$$C_l = \frac{1}{2}\sum_{n=1}^{2N}\{h_{n-1}\cos[(n-1-N)\sigma_l\Delta T] + h_n\cos[(n-N)\sigma_l\Delta T]\}, \quad (5)$$
$$l = 0,1,\cdots,M,$$

$$S_l = \frac{1}{2}\sum_{n=1}^{2N}\{h_{n-1}\sin[(n-1-N)\sigma_l\Delta T] + h_n\sin[(n-N)\sigma_l\Delta T]\}, \quad (6)$$
$$l = 1,2,\cdots,M.$$

Therefore Eq. (4) can be decomposed into two linear equations

$$FX = C, \quad GY = S. \quad (7)$$

The accuracy of tidal prediction can be improved as longer data time series are analyzed and more tidal constituents are selected in Eq. (7).

**ITERATING ALGORITHM FOR COMPUTING $C$ AND $S$**

In terms of complex form

$$C_l + iS_l = \frac{1}{2}\sum_{n=1}^{2N}\left\{h_{n-1}e^{i(n-1-N)\sigma_l\Delta T} + h_n e^{i(n-N)\sigma_l\Delta T}\right\} \quad l = 1,2,\cdots,M. \quad (8)$$

Using Goertzel iteration [11],

$$\begin{cases} \rho_k = h_k + 2\rho_{k+1}\cos(\sigma_l \Delta T) - \rho_{k+2} & k = 2N-1, 2N-2, \cdots, 1 \\ \rho_0 = \dfrac{h_0}{2} + 2\rho_1\cos(\sigma_l \Delta T) - \rho_2, & \end{cases} \quad (9)$$

under initial conditions

$$\rho_{2N+1} = 0, \quad \rho_{2N} = \frac{h_{2N}}{2}.$$

After $2N$ time iterations, whence

$$C_l + iS_l = \left(\rho_0 - \rho_1 e^{-i\sigma_l \Delta T}\right) e^{-iN\sigma_l \Delta T}. \quad (10)$$

In this method, only $2N$ multiplications are needed.

## HARMONIC METHOD FOR SEGMENTS OF REGULAR OBSERVATIONS

For $K$ equal-length segments of observed data (overlapping is allowed). $h_n^{(k)}$ ($n = 0, 1, \cdots, 2N$, $k = 1, 2, \cdots, K$) are observations starting at time $t_0^{(k)}$ in $k$th segment. The corresponding random errors are $r_n^{(k)}$. For each segment, the same $M$ is chosen. The corresponding $R$ and $H$ have

$$\overline{R}_k = \left(\frac{r_0^{(k)}}{\sqrt{2}}, r_1^{(k)}, \cdots, r_{N-1}^{(k)}, \frac{r_N^{(k)}}{\sqrt{2}}\right)^T, \overline{H}_k = \left(\frac{h_0^{(k)}}{\sqrt{2}}, h_1^{(k)}, \cdots, h_{N-1}^{(k)}, \frac{h_N^{(k)}}{\sqrt{2}}\right)^T$$
$$k = 1, 2, \cdots, K.$$

Then

$$\overline{R} = \overline{H} - \overline{Q} W Z, \quad (11)$$

where

$$\overline{R}^T = (\overline{R}_1^T, \overline{R}_2^T, \cdots, \overline{R}_K^T), \quad \overline{H}^T = (\overline{H}_1^T, \overline{H}_2^T, \cdots, \overline{H}_K^T).$$

For $\overline{Z}^T = (\overline{X}^T, \overline{Y}^T)$, the components of $\overline{X}$ and $\overline{Y}$ are

$$\overline{x}_0^{(k)} = x_0^{(k)},$$

$$\left\{\overline{x}_m^{(k)}, \overline{y}_m^{(k)}\right\} = R_m^{(k)}\left\{\cos(\varphi_m^{(k)} - N\sigma_m \Delta T), \sin(\varphi_m^{(k)} - N\sigma_m \Delta T)\right\} \quad (12)$$
$$m = 1, 2, \cdots, M.$$

$$\overline{Q} = \begin{bmatrix} Q & & & \\ & Q & & \\ & & \ddots & \\ & & & Q \end{bmatrix}$$

is a $[(2N+1)K] \times [(2M+1)K]$ matrix.

$$W^T = [W_1, W_2, \cdots, W_K], \quad (13)$$

where

$$W_k = \begin{bmatrix} U_1^{(k)} & U_3^{(k)^T} \\ -U_3^{(k)} & U_2^{(k)} \end{bmatrix}, \quad k = 1, 2, \cdots, K,$$

and

$$U_1^{(k)} = \text{diag}\left[1, \cos(t_0^{(k)}\sigma_1), \cdots, \cos(t_0^{(k)}\sigma_M)\right],$$

$$U_2^{(k)} = \text{diag}\left[\cos(t_0^{(k)}\sigma_1), \cos(t_0^{(k)}\sigma_2), \cdots, \cos(t_0^{(k)}\sigma_M)\right],$$

$$U_3^{(k)} = \begin{bmatrix} 0 & \sin(t_0^{(k)}\sigma_1) & & & \\ 0 & & \sin(t_0^{(k)}\sigma_2) & & \\ \vdots & & & \ddots & \\ 0 & & & & \sin(t_0^{(k)}\sigma_M) \end{bmatrix}.$$

From

$$\|\overline{R}\|_2^2 = \|\overline{H} - \overline{Q} W \overline{Z}\|_2^2 = \min, \quad (14)$$

we have the normal equations

$$A\overline{Z} = P, \quad (15)$$

where

$$A = \sum_{k=1}^{K} \begin{bmatrix} U_1^{(k)} F U_1^{(k)} + U_3^{(k)^T} G U_3^{(k)} & U_1^{(k)} F U_3^{(k)^T} - U_3^{(k)^T} G U_2^{(k)} \\ U_3^{(k)} F U_1^{(k)} - U_2^{(k)} G U_3^{(k)} & U_3^{(k)} F U_3^{(k)^T} + U_2^{(k)} G U_2^{(k)} \end{bmatrix},$$

and $P^T = \left(\overline{C}^T, \overline{S}^T\right)$. The components of $\overline{C}$ and $\overline{S}$ are

$$\overline{C}_l = \frac{1}{2} \sum_{k=1}^{K} \sum_{n=1}^{2N} \left\{ h_{n-1}^{(k)} \cos\left[\left(n - 1 - N + \frac{t_0^{(k)}}{\Delta T}\right)\sigma_l \Delta T\right] \right.$$
$$\left. + h_n^{(k)} \cos\left[\left(n - N + \frac{t_0^{(k)}}{\Delta T}\right)\sigma_l \Delta T\right] \right\} \quad l = 0, 1, \cdots, M,$$

$$\overline{S}_l = \frac{1}{2} \sum_{k=1}^{K} \sum_{n=1}^{2N} \left\{ h_{n-1}^{(k)} \sin\left[\left(n - 1 - N + \frac{t_0^{(k)}}{\Delta T}\right)\sigma_l \Delta T\right] \right.$$
$$\left. + h_n^{(k)} \sin\left[\left(n - N + \frac{t_0^{(k)}}{\Delta T}\right)\sigma_l \Delta T\right] \right\} \quad l = 1, 2, \cdots, M.$$

## CONSTRAINED CONDITIONS FOR INADEQUATE OBSERVED DATA

In the circumstances of analyzed data with insufficient-length (manly tidal currents), the tidal constituents can not be separated effectively due to ill conditioning appeared in Eqs. (7) and (15). Some constrained conditions must be provided.

For two tidal constituents, we have the following relationship proposed by Dronkers [12]

$$F_{2j-1}: \bar{x}_{2j-1} - \alpha_{2j-1,2j}(\bar{x}_{2j}\cos\theta_{2j-1,2j} + \bar{y}_{2j}\sin\theta_{2j-1,2j}) = 0,$$
$$j = 1, 2, \cdots, J, \quad (16)$$
$$F_{2j}: \bar{y}_{2j-1} - \alpha_{2j-1,2j}(\bar{y}_{2j}\cos\theta_{2j-1,2j} - \bar{x}_{2j}\sin\theta_{2j-1,2j}) = 0,$$

where $J$ is the number of couples of tidal constituents to be chosen among the $M$ tidal constituents. $\alpha_{2j-1,2j}$ and $\theta_{2j-1,2j}$ are functions of $j$. In matrix notation, Eq. (16) can be expressed as

$$B\bar{Z} = 0, \quad (17)$$

$$B = \begin{bmatrix} D_1 & & & \overbrace{0 \; 0 \; \cdots \; 0}^{(M+1)K-2J} & E_1 & & & \overbrace{0 \; 0 \; \cdots \; 0}^{MK-2J} \\ & D_2 & & 0 \; 0 \; \cdots \; 0 & & E_2 & & 0 \; 0 \; \cdots \; 0 \\ & & \ddots & \vdots\;\vdots\;\cdots\;\vdots & & & \ddots & \vdots\;\vdots\;\cdots\;\vdots \\ & & & D_J \; 0 \; 0 \; \cdots \; 0 & & & & E_J \; 0 \; 0 \; \cdots \; 0 \end{bmatrix},$$

where

$$D_j = \begin{bmatrix} 1 & -\alpha_{2j-1,2j}\cos\theta_{2j-1,2j} \\ 0 & \alpha_{2j-1,2j}\sin\theta_{2j-1,2j} \end{bmatrix}, \; E_j = \begin{bmatrix} 0 & -\alpha_{2j-1,2j}\sin\theta_{2j-1,2j} \\ 1 & -\alpha_{2j-1,2j}\cos\theta_{2j-1,2j} \end{bmatrix},$$
$$j = 1, 2, \cdots, J.$$

If taking

$$\|\bar{R}\|_2^2 + \sum_{j=1}^{2J} \lambda_j F_j = \min,$$

we can get the linear equations

$$\begin{bmatrix} A & B^T \\ B & O \end{bmatrix} \begin{bmatrix} \bar{Z} \\ \Lambda \end{bmatrix} = \begin{bmatrix} P \\ O \end{bmatrix}, \quad (18)$$

where the vector of Lagrange multiplier is

$$\Lambda = (\lambda_1, \lambda_2, \cdots, \lambda_{2J})^T.$$

In order to eliminate unknown $\Lambda$ in Eq. (18), we take

$$B^* = \begin{bmatrix} \overbrace{D_1^* \qquad\qquad\qquad}^{(M+1)K} & \overbrace{E_1^* \qquad\qquad\qquad}^{MK} \\ \;\; D_2^* & \;\; E_2^* \\ \;\;\;\; \ddots & \;\;\;\; \ddots \\ \;\;\;\;\;\; D_J^* & \;\;\;\;\;\; E_J^* \\ \;\;\;\;\;\;\;\; 1 & \\ \;\;\;\;\;\;\;\;\;\; 1 & \\ \;\;\;\;\;\;\;\;\;\;\;\; \ddots & \\ \;\;\;\;\;\;\;\;\;\;\;\;\;\; 1 & \\ & \;\;\;\;\;\;\;\;\;\;\;\;\;\;\;\; 1 \\ & \;\;\;\;\;\;\;\;\;\;\;\;\;\;\;\;\;\; 1 \\ & \;\;\;\;\;\;\;\;\;\;\;\;\;\;\;\;\;\;\;\; \ddots \\ & \;\;\;\;\;\;\;\;\;\;\;\;\;\;\;\;\;\;\;\;\;\; 1 \end{bmatrix},$$

where

$$D_j^* = \begin{bmatrix} 1 & \alpha_{2j-1,2j}^{-1}\cos\theta_{2j-1,2j} \\ 0 & -\alpha_{2j-1,2j}^{-1}\sin\theta_{2j-1,2j} \end{bmatrix}, \; E_j^* = \begin{bmatrix} 0 & \alpha_{2j-1,2j}^{-1}\sin\theta_{2j-1,2j} \\ 1 & \alpha_{2j-1,2j}^{-1}\cos\theta_{2j-1,2j} \end{bmatrix},$$
$$j = 1, 2, \cdots, J.$$

Taking advantage of $B^* B^T = O$, we have

$$\begin{bmatrix} B^* & O \\ O & I \end{bmatrix} \begin{bmatrix} A & B^T \\ B & O \end{bmatrix} \begin{bmatrix} \bar{Z} \\ \Lambda \end{bmatrix} = \begin{bmatrix} B^* P \\ O \end{bmatrix},$$

where $I$ is unit matrix. Now the rank of matrix equation Eq. (18) is reduced by an equivalent form

$$\begin{bmatrix} B^* A \\ B \end{bmatrix} \bar{Z} = \begin{bmatrix} B^* P \\ O \end{bmatrix}. \quad (19)$$

## CONCLUDING REMARKS

The least squares method has broadly been adopted in tidal harmonic analysis. For a concrete problem using a computer, a better algorithm not only requests a less computing time but also is able to identify effectively the tidal constituents from observed data. An efficient algorithm of tidal harmonic analysis and prediction is presented here. It is the algorithm that can calculate coefficients of normal equations very simply and efficiently. To compute the right-hand terms of the normal equations, Goertzel iteration [11] is adopted to accomplish whole calculation processes quickly and accurately. In order to handle the segments of the observed date (mainly adapted to analyze tidal currents), a general algorithm for $K$ sets of real-time regularly observed date in equal observing length can be derived from above results.

If the above algorithm is used to analyze the tidal constituents, the total analyzed data must have sufficient length. Otherwise ill conditioning in the system of equations appears so that conventional algorithm can not separate tidal constituents effectively. Consequently in the circumstances of insufficient data (manly tidal currents), some constrained conditions can be established based on known approximate relationships among the harmonic constants of the tidal constituents. Then the lease squares solutions can be obtained from these constrained conditions. To various circumstances, the resultant linear equations can be deduced from this algorithm in order to avoid appropriately the emergence of ill conditioning. Because the constrained conditions can reasonably be established, usually final solutions do not have to be corrected so that the harmonic constants can be determined with sufficient accuracy.

**REFERENCES**

1. Newton, I., 1687, *Philosophia Naturalis Principia Mathematica*.
2. Thomson, W., 1868-1876, "Reports of committee for harmonic analysis," *British Association for the Advancement of Science*.
3. Darwin, G. H., 1883-1886, "Reports of a committee for the harmonic analysis of tides," *British Association for the Advancement of Science*.
4. Harris, R. A., 1897-1907, "Manual of tides," *Appendices to Reports of the U.S. Coast and Geodetic Survey*.
5. Doodson, A. T., 1921, "The harmonic development of the tide-generating potential," *Proceedings of the Royal Society of London Series A. Mathematical and Physical Sciences*, Vol. **100**, pp. 305-329.
6. Amin, M., 1976, "The fine resolution of tidal harmonics," *Geophysical Journal of the Royal Astronomical Society*, Vol. **44**, No. 2, pp. 293-310.
7. Amin, M., 1987, "A method for approximating the nodal modulations of the real tide," *International Hydrographic Review*, Vol. **64**, No. 2, pp. 103-113.
8. Amin, M., 1991, "Superfine resolution of tidal harmonic constants," in: *Tidal Hydrodynamics*, BB Parker, ed., John Wiley & Sons, Inc., pp. 711-724.
9. George, K. J., and Simon, B., 1984, "The species concordance method of tide prediction in estuaries," *International Hydrographic Review*, Vol. **65**, No. 1, pp. 121-146.
10. Simon, B., 1991, "The species concordance method of tide prediction," in: *Tidal Hydrodynamics*, BB Parker, ed., John Wiley & Sons, Inc., pp. 725-735.
11. Goertzel, G., 1958, "An algorithm for the evaluation of finite trigonometric series," *The American Mathematical Monthly*, Vol. **65**, No. 1, pp. 34-35.
12. Dronkers, J. J., 1964, *Tidal computations in rivers and coastal waters*, Amsterdam: North-Holland Publishing Co.